\documentclass{aa}
\usepackage{graphicx}
\usepackage{txfonts}
\usepackage{natbib}
\usepackage{hyperref}
\newcommand{\mearth}{M_\oplus}

\def\ms{\hbox{\,m\,s$^{-1}$}}         
\def\m2s2{\hbox{\,m$^{2}$\,s$^{-2}$}} 
\def\sini{\hbox{sin\,$i$}}      
\def\Msun{\hbox{$\mathrm{M}_{\odot}$}}             

\begin{document} 

\title{Proxima Centauri reloaded: Unravelling the stellar noise in radial velocities}


   \author{M. Damasso
          \inst{1} $\&$
          F. Del Sordo
          \inst{2,3}
          }

   \institute{INAF-Osservatorio Astrofisico di Torino, via Osservatorio 20, 10025, Pino Torinese, Italy\\
              \email{damasso@oato.inaf.it}
         \and
             Department of Geology \& Geophysics, and Department of Astronomy, Yale University, 06511 New Haven, CT, USA
             \and 
             Nordita, Royal Institute of Technology and Stockholm University, SE-10691 Stockholm, Sweden \\
             \email{fabio.delsordo@yale.edu}
             }

   \date{Received - - -; Accepted, - - - }

 
  \abstract
  {The detection and characterisation of Earth-like planets with Doppler signals of the order of 1 \ms currently represent one of the greatest challenge for extrasolar-planet hunters. As results for such findings are often controversial, it is desirable to provide independent confirmations of the discoveries.  Testing different models for the suppression of non-Keplerian stellar signals usually  plaguing radial velocity data is essential to ensuring findings are robust and reproducible.}
   {Using an alternative treatment of the stellar noise to that discussed in the discovery paper, we re-analyse the radial velocity dataset that led to the detection of a candidate terrestrial planet orbiting the star Proxima Centauri. We aim to confirm the existence of this outstanding planet, and  test the existence of a second planetary signal.}
   {Our technique jointly modelled Keplerian signals and residual correlated signals in radial velocities using Gaussian processes. We analysed only radial velocity measurements without including other ancillary data in the fitting procedure. In a second step, we have compared our outputs with results coming from photometry, to provide a consistent physical interpretation. Our analysis was performed in a Bayesian framework to quantify the robustness of our findings.}
   {We show that the correlated noise  can be successfully modelled as a Gaussian process regression, and contains a periodic term modulated on the stellar rotation period and characterised by an evolutionary timescale of the order of one year. Both findings appear to be robust when compared with results obtained from archival photometry, thus providing a reliable description of the noise properties. We confirm the existence of a coherent signal described by a Keplerian orbit equation that can be attributed to the planet Proxima\,b, and provide an independent estimate of the planetary parameters. Our Bayesian analysis dismisses the existence of a second planetary signal in the present dataset.}
   {}

   \titlerunning{Uravelling stellar noise in Proxima Centauri radial velocities}
   \maketitle
%

\section{Introduction}
The discovery of a potentially terrestrial planet in an approximately 11-day orbit around the M dwarf Proxima Centauri, announced by \cite{anglada16} (hereafter AE16), represents a major breakthrough in the extrasolar-planet science. Proxima is the closest known star to the Sun, and the planet Proxima\,b might have an equilibrium surface temperature suitable for liquid water to exist. 
Intrinsic stellar activity can mask planetary signals measured in radial
velocity time series, especially for M dwarf systems. This stellar
activity signal can be of the order of a few \ms, larger than the semi-amplitude of the planetary
signal previously measured for the Proxima system.
To date, several techniques have been developed to mitigate this effect. \cite{dumusque16b} devised a blind test, based on simulated RV measurements including realistic stellar activity noise \citep{dumusque16a}, to compare the performances of different methods. To derive orbital and physical properties of Proxima\,b from high-precision RV measurements it is therefore necessary a proper treatment of the stellar activity (the 'noise') contribution. To do so, AE16 used a Bayesian framework consisting of a moving average term and linear correlations with activity indexes.
The technique used by AE16 was tested on the RV datasets of \cite{dumusque16b} and found to be very effective for recovering signals induced by low-mass planets. A different Bayesian technique applied to the test which performed similarly well is based on modelling the correlated noise using Gaussian processes (GPs) (see e.g. \citealt{rasmu06} and \citealt{roberts12} for a general description of GPs). GPs are a powerful tool to mitigate the correlated noise in a set of measurements, such as the stellar activity signature in RV data. This technique has been used  successfully in several recent works dealing with real RV data (\citealt{hay14}, \citealt{grun15}, \citealt{rajpaul15}, \citealt{affer16}, \citealt{faria16}, \citealt{morales16}).
In this article we have applied the GP technique to the radial velocity observations of Proxima published in AE16, taken with the UVES and HARPS spectrographs. HARPS data are divided in a pre-2016 and 2016 (PRD: Pale Red Dot campaign) subsamples. We aim to confirm and characterise the signal of Proxima\,b in an independent way through an alternative and robust modelling of the stellar noise. We also consider the existence of a second planetary signal, by testing the hypothesis in a Bayesian framework.


\section{Description of the model and technique}
        \label{sec:modelde}
We model the correlated noise by adopting a quasi-periodic covariance function, similarly to what was done, for example, in \cite{affer16}. This function is described by parameters, called hyperparameters, which try to model in a simple fashion some of the physical phenomena undelying the stellar noise.
The quasi-periodic kernel is described by the covariance matrix
\begin{eqnarray} \label{eq:eqgpkernel}
k(t, t^{\prime}) = h^2\cdot\exp\bigg[-\frac{(t-t^{\prime})^2}{2\lambda^2} - \frac{sin^{2}(\dfrac{\pi(t-t^{\prime})}{\theta})}{2w^2}\bigg] + \nonumber \\
+\, (\sigma^{2}_{\rm instr, RV}(t)\,+\,\sigma_{\rm inst,jit}^{2})\cdot\delta_{\rm t, t^{\prime}}
,\end{eqnarray}
where $t$ and $t^{\prime}$ indicate two different epochs. 
This kernel is composed by a periodic term coupled to an exponential decay one, in order to model a recurrent signal linked to stellar rotation and taking into account the size-evolution of finite-lifetime active regions. Such approach is therefore particularly suitable to model the stellar noise on short-term timescales, as we consider that modulated by the stellar rotation period.
The hyperparameter $h$ of the covariance function represents the amplitude of the correlations; $\theta$ is related to the rotation period of the star; $w$ is the length scale of the periodic component, linked to the size evolution of the active regions; $\lambda$ is the correlation decay timescale, that we assume to be related to the active regions lifetime.
Here, $\sigma_{\rm instr, RV}(t)$ is the RV internal error at time \textit{t} for a given instrument; $\sigma_{\rm instr, jit}$ is the additional uncorrelated 'jitter' term, one for each instrument, that we add in quadrature to the internal errors to take into account instrumental effects and other $\lesssim 1 \ms$ noise sources included neither in  $\sigma_{\rm instr, RV}(t)$ nor in our stellar activity framework; $\delta_{\rm t, t^{\prime}}$ is the Dirac delta function.

In the GP framework, the log-likelihood function to be maximised by the Markov chain Monte Carlo (MCMC) procedure is 
\begin{equation} \label{eq:2-1}
\ln \mathcal{L} = -\frac{n}{2}\ln(2\pi) - \frac{1}{2}\ln(det\,\mathbf{K}) - \frac{1}{2}\overline{r}^{T}\cdot\mathbf{K}^{-1}\cdot\overline{r}
,\end{equation}
where $n$ is the number of the data points, \textbf{K} is the covariance matrix built from the covariance function in Equation (\ref{eq:eqgpkernel}), and $\overline{r}$ represents the RV residuals, obtained by subtracting the Keplerian signal(s) from the original RV dataset. 

The general form for the models that we tested in this work is given by the equation
\begin{eqnarray}\label{eq:eqrvmodel}
\Delta RV(t) =  \gamma_{\rm instr} + \sum_{\rm j=1}^{n_{planet}} \Delta RV_{\rm Kep,j}(t)
                   +\, \dot{\Delta RV(t)} +
\nonumber \\
                        +\, \Delta RV(t)_{\rm(activity,\, short-term)} =
\nonumber \\
= \gamma_{\rm instr.} \hspace{-.05cm} + \hspace{-.05cm} \sum_{\rm j=1}^{n_{planet}}\hspace{-.05cm} K_{\rm j}\large[\cos(\nu_{\rm j}(t, e_{\rm j}, T_{\rm c,j}, \textit{P}_{\rm j}) + \omega_{\rm j}) \hspace{-.05cm} +  e_{\rm j}\cos(\omega_{\rm j})\large]  \hspace{-.05cm} + 
\nonumber \\
              + \dot{\Delta RV(t)} + \hspace{-.05cm} GP
,\end{eqnarray}
where $n_{\rm planet}=1,2$; $\nu$ is function of time \textit{t}, time of the inferior conjuntion $T_{\rm c,j}$, orbital period $\textit{P}_{\rm j}$, eccenticity \textit{e} and argument of periapsis $\omega_{\rm j}$; $\gamma_{\rm instr}$ is the RV offset, one for each instrument; $\dot{\Delta RV(t)}$ is the secular acceleration; $GP$ is the stellar noise modelled with the Gaussian process.
Instead of fitting separately $e_{\rm j}$ and $\omega_{\rm j}$, we use the auxiliary parameters
$C_{\rm j} = \sqrt{e_{\rm j}}\cdot \cos \omega_{\rm j}$ and $S_{\rm j} = \sqrt{e_{\rm j}} \cdot \sin \omega_{\rm j}$ to reduce the covariance between $e_{\rm j}$ and $\omega_{\rm j}$.

Our analysis is performed in two steps through two separate MCMC runs.
First, we investigate the existence of the signal attributed to Proxima\,b by including one Keplerian in the model together with the GP noise term. 
Through our GP analysis, we can constrain the
values of our model hyperparameters, and thus constrain the stellar
properties we assume these hyperparameters are related to.
Here we examine the existence of a second planetary signal in the RV data, whose presence could not be ruled out at long orbital periods, as suggested by AE16. Finally, we evaluate the Bayesian evidences (marginal likelihoods) for the two models to assess whether the present RV dataset supports the two-planet scenario over the other. 
We use the publicly available \texttt{emcee} algorithm \citep{foreman13} to perform the MCMC analyses of the RV data, and the publicly available \texttt{GEORGE} Python library to perform the GP fitting within the MCMC framework \citep{ambi14}. We used 150 random walkers for each MCMC run, and to derive the parameter posterior distributions we performed a burn-in by removing from the samples those with $\ln \mathcal{L}$ lower than the median of the whole $\ln \mathcal{L}$ dataset. 
The strategy of our MCMC analysis is to put as little constraint as possible to the model parameters. Following such a philosophy, the starting points of the walkers in the parameter space were randomly chosen from normal distributions with $\sigma$ sufficiently large to allow a broad exploration of the parameter space around guess values. We also used uninformative priors for almost all the jump parameters. The MCMC convergence is checked using the diagnostic proposed by \cite{ford06}, and the best fit values and uncertainties for each jump parameter are calculated as the median of the marginal posterior distributions and the 16$\%$ and 84$\%$ quantiles.
We quantify the relative statistical significance of the two tested models by deriving their Bayesian evidences  $\mathcal{Z}_{1}$ and $\mathcal{Z}_{2}$, according to Bayes' theorem. The selection of the model better describing the data is made by calculating the ratio $\mathcal{Z}_{1}$/$\mathcal{Z}_{2}$. In fact, by assuming that there is not any a priori reason to favour one model over the other, the $\mathcal{Z}$-ratio represents the Bayes factor, that is, the figure of merit encoding all the support the data give to one model over the other (see, e.g. \cite{diaz16} for a detailed discussion in the framework of the extrasolar-planets detection). We estimate the natural logarithm of each evidence $\mathcal{Z}_{i}$ by using two analytic approximations due to \cite{chib01} and \cite{perra14} (hereafter indicated as the CJ01 and Perr14 estimators). For the statistical interpretation we follow here a conventional empirical scale, the so-called Jeffreys' scale, which states that the model with the highest $\mathcal{Z}$ is strongly favoured over the other when $\Delta\ln\mathcal{Z}\geqslant5$, while $2.5<\Delta\ln\mathcal{Z}<5$ denotes moderate evidence, $1<\Delta\ln\mathcal{Z}<2.5$ weak evidence, and $\Delta\ln\mathcal{Z}<1$ corresponds to inconclusive evidence. 
\begin{table}
  \caption[]{Prior probability distributions for the one-planet model parameters.}
         \label{priors1plan}
         \scriptsize
   \begin{tabular}{l l l l}
            \hline
            \noalign{\smallskip}
            Jump parameter     &  Prior & Lower bound & Upper bound \\
            \noalign{\smallskip}
            \hline
            \noalign{\smallskip}
            h [\ms] & Uniform & 0 & 10\\
            $\lambda$ [days] & Uniform & 0 & 6000            \\
            $w$ & Uniform & 0 & 100\\
            $\ln\theta$ [days] & Uniform & 0 ($\theta$=1) & $\ln(120)$\\
            $K_{\rm 1}$ [\ms] & Uniform & 1 & 10 \\
            $\ln P_{\rm1}$ [days] & Uniform & 0 (P$_{1}$=1) & $\ln6000$\\
            $T_{\rm 1,c}$ [JD-2\,400\,000] & Uniform & 51500 & 57500 \\
            $C_{\rm 1}=\sqrt[]{e_{\rm 1}}\cdot\cos\omega_{\rm 1}$ & Uniform & -1 & 1 \\
            $S_{\rm 1}=\sqrt[]{e_{\rm 1}}\cdot\sin\omega_{\rm 1}$ & Uniform & -1 & 1 \\
            $e_{\rm 1}$ (=$C_{\rm 1}^{2}$+$S_{\rm 1}^{2}$) & 3.1$\cdot(1-e_{\rm 1})^{2.1}$ & 0 & 1\\
            $\omega_{\rm 1}$ [rad] & Uniform & 0 & 2$\pi$\\
            $dV_{\rm r}/dt$ [\ms\,day$^{-1}$] & Uniform & -0.001 & +0.001 \\
            $\gamma_{\rm HARPS_{pre-2016}}$ [\ms] & Uniform & -3 & +3 \\
            $\gamma_{\rm HARPS_{PRD}}$ [\ms] & Uniform & -3 & +3 \\
            $\gamma_{\rm UVES}$ [\ms] & Uniform & -3 & +3 \\
            $\sigma_{\rm jit, HARPS_{pre-2016}}$ [\ms] & Uniform & 0 & 10 \\ 
            $\sigma_{\rm jit, HARPS_{PRD}}$ [\ms] & Uniform & 0 & 10 \\ 
            $\sigma_{\rm jit, UVES}$ [\ms] & Uniform & 0 & 10 \\ 
            \noalign{\smallskip}
            \hline
     \end{tabular}    
   \end{table}

\begin{figure*}[!htbp]
   \caption{ Distributions of total MCMC values for the jump parameters of the one-planet model as function of the natural logarithm of the likelihood function.}
   \centering
   \includegraphics[width=19cm]{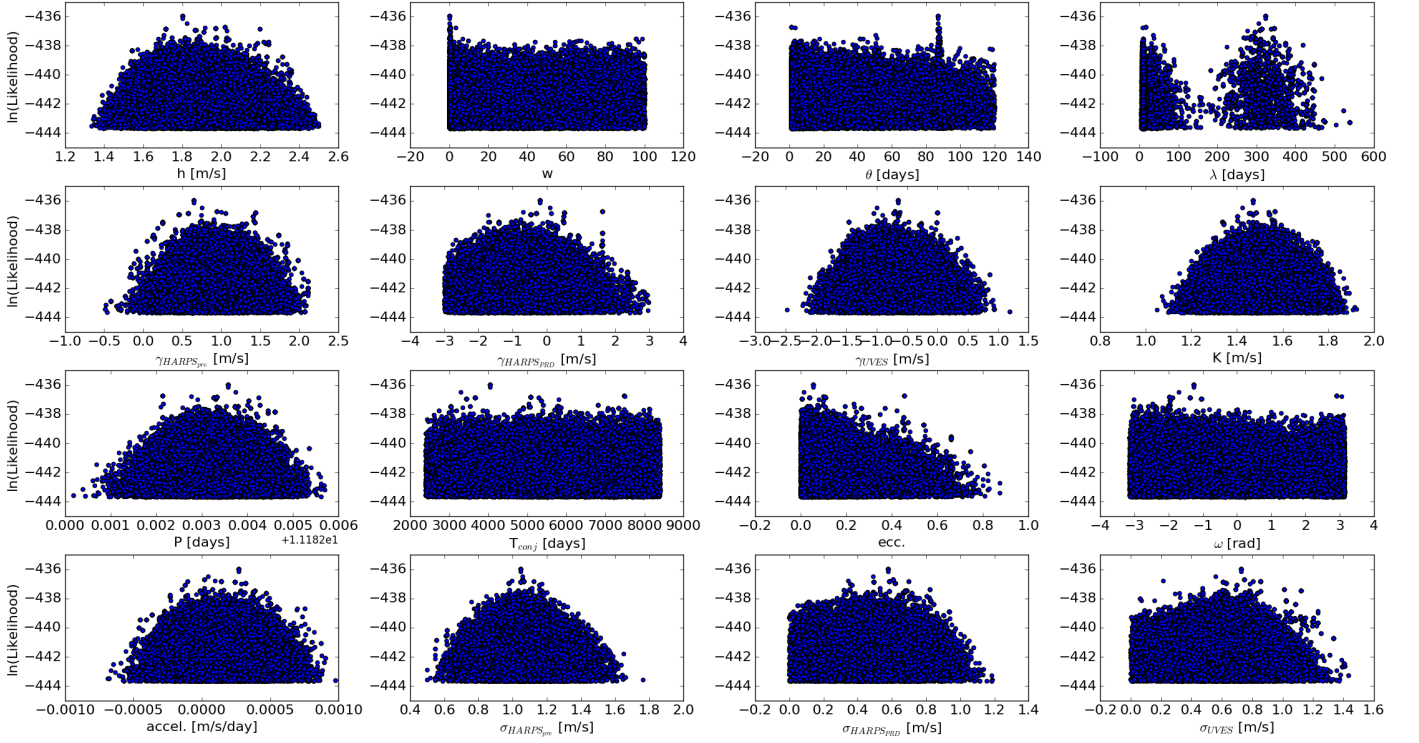}
              \label{lnLikevsparam1planet_full}%
    \end{figure*}

\begin{figure*}[!htbp]
   \caption{ Distributions 
   as in Fig. \ref{lnLikevsparam1planet_full} after constraining the GP hyperparameter $\theta$ in the range [85, 89] days. 
The vertical red lines represent the median values (solid) and the 16th and 84th percentiles (dotted) of these distributions, which are listed in Table \ref{percentilesoneplan}.}
   \centering
   \includegraphics[width=19cm]{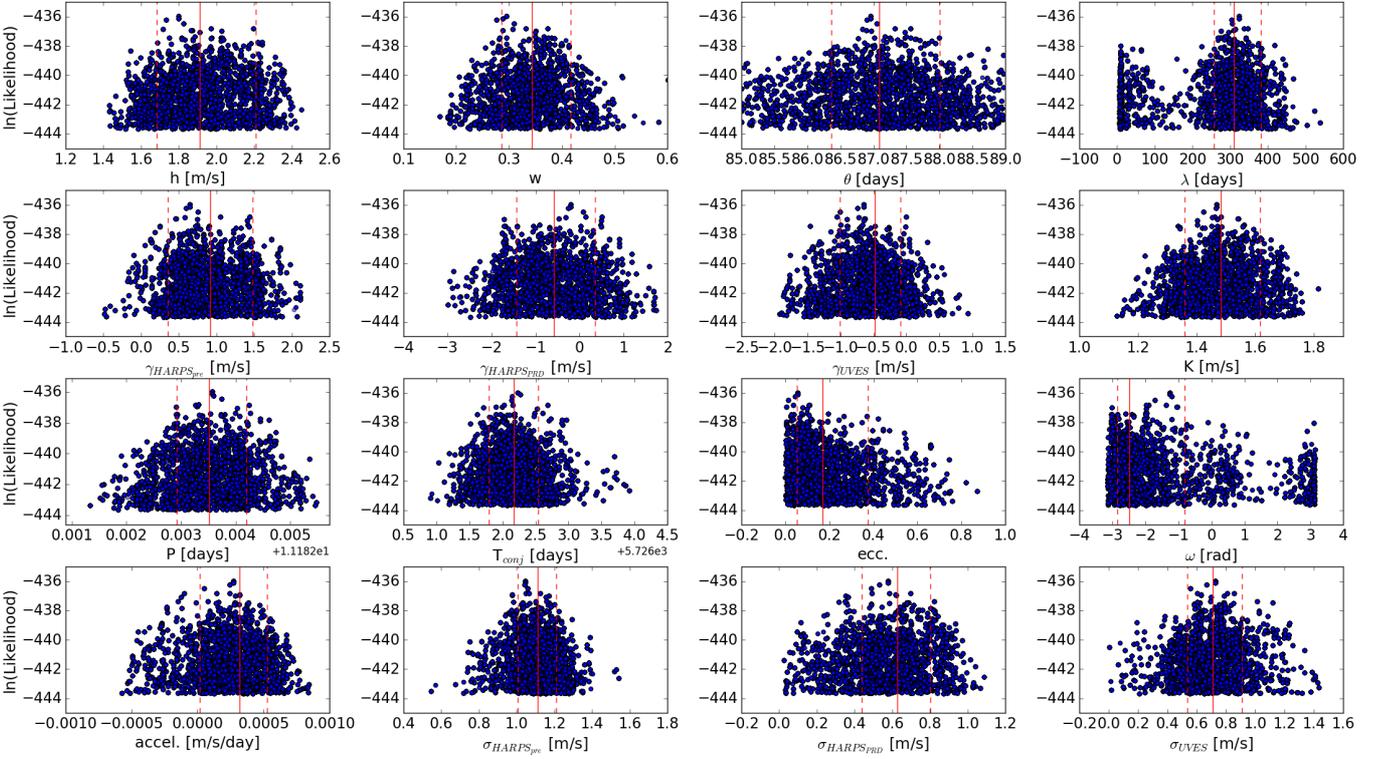}
              \label{lnLikevsparam1planet_theta}%
    \end{figure*}

\section{One-planet model}
\label{sec:onepl}
\begin{figure}
   \caption{ Radial velocity residuals, after subtracting the best-fit GP correlated noise model, folded at the best-fit orbital period $P$=11.1855 days. Different symbols and colours are used for each dataset. Grey dots represent average values for 20 phase bins. Superposed are the best-fit eccentric model (continuous red curve) and the simpler circular model (dashed red curve).}
   \centering 
   \includegraphics[width=10cm]{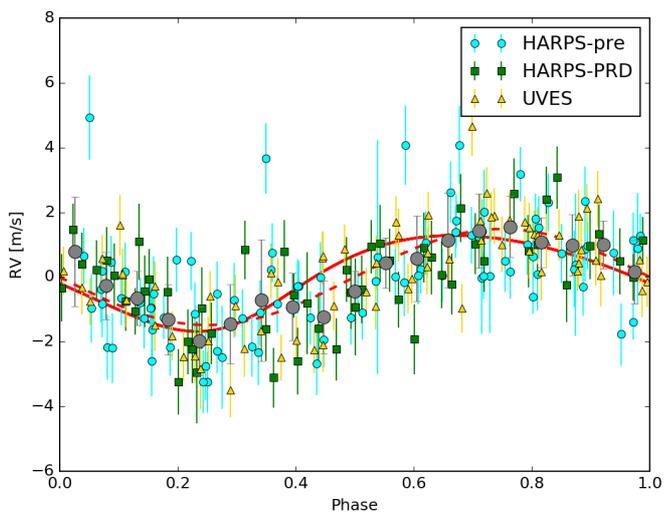}
              \label{foldedrv1}%
\end{figure}

\begin{figure}[!htbp]
   \caption{ Radial velocity residuals for the one-planet model, after subtracting the best-fit GP correlated noise model, folded at the best-fit orbital period $P$=11.1855 days. Each dataset is showed here separately for clarity. Superposed are the best-fit eccentric model (continuous red curve) and the simpler circular model (dashed red curve).}
   \includegraphics[width=10cm]{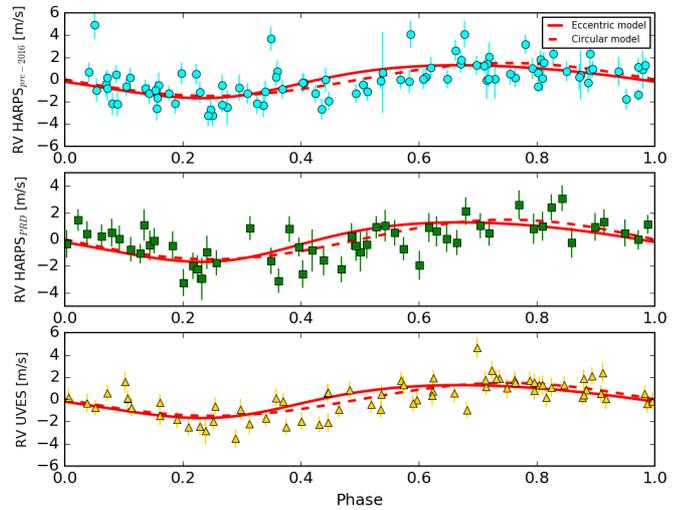}
              \label{foldedrv}%
\end{figure}

\begin{figure}[!htbp]
   \caption{ Radial velocity residuals time series (black dots), after subtracting our best-fit orbital solution for Proxima\,b. The blue line with grey shaded 1-$\sigma$ regions represents our best-fit GP quasi-periodic model for the correlated stellar noise. The upper plot shows the complete dataset, while the two plots in the second row show selected epochs, to easier visualize the agreement between the model and the data.}
     \centering 
   \includegraphics[width=10cm]{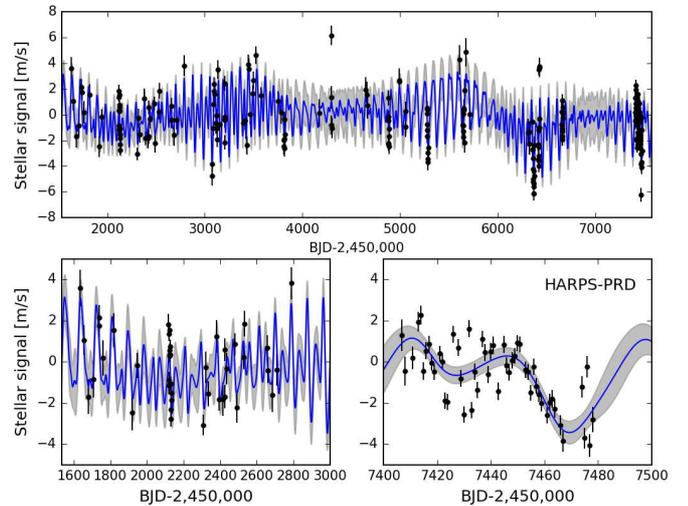}
              \label{noisemodel}%
\end{figure}

In order to minimally constrain our
model, we have chosen Bayesian prior probability distributions as described in Table \ref{priors1plan}. All are uniform (uninformative) priors, except for the orbital eccentricity: for this we adopted a Beta distribution, because it is well-known that the noise tends otherwise to favour higher eccentricities
(e.g. see Appendix B.4 in \citealt{gregory10}, April 2016 supplement\footnote{Available at http://www.cambridge.org/pl/academic/subjects/statistics-probability/statistics-physical-sciences-and-engineering/bayesian-logical-data-analysis-physical-sciences-comparative-approach-mathematica-support}). We considered the GP periodic hyperparameter $\theta$ and the orbital period \textit{P} as scale invariant parameters, for which an uninformative prior is one that is uniform in $\ln P$. We have chosen a very large range for the GP hyperparameter \textit{w}, which usually is constrained to values less than one when there is clear evidence of a stellar-rotation-linked periodic component in the correlated noise (e.g. \citealt{morales16}).  
Our MCMC run stopped at $\sim$200\,000 steps: the median of $\ln \mathcal{L}$ calculated over the 150 chains maintained a nearly constant value over $\sim$140\,000 steps. After the burn-in of the first 100\,000 steps, we have performed an additional burn-in in the $\ln \mathcal{L}$ space by deriving the marginal posterior distributions of samples with $\ln \mathcal{L}$ > median($\ln \mathcal{L}$) (see Fig. \ref{lnLikevsparam1planet_full}). The posterior distribution of the GP hyperparameter $\theta$ shows a concentrated and symmetrical sub-sample between 85 and 89 days with the highest values of $\ln \mathcal{L}$, which is the range where the known rotation period of Proxima falls in. Considering that $\theta$ is expected to represent the stellar rotation period, this evidence is particularly relevant because it reveals that the correlated noise  does contain a periodic term which is modulated by the rotation of the star. The rotation period of Proxima has been robustly estimated from ASAS \textit{V}-band photometry over an interval spanning up to 15 years (e.g. see Fig. 3 in \citealt{robertson16}, or Fig. 3 in \citealt{wargelin16}). In particular, \cite{wargelin16} present the Lomb-Scargle periodogram of the ASAS-3 and ASAS-4 photometry covering 15 years, which shows a very significant and sharp peak at 83.1 days. The same authors show evidence for differential rotation in the light curve by analysing periodograms of different observing seasons, finding significant peaks ranging from \textit{P}$_{\rm rot}$=77 to \textit{P}$_{\rm rot}$=90 days. Based on this robust result from photometry, we can naturally restrict our analysis to the 697\,510 posterior samples for which $\theta$ falls in the range 85-89 days, 
ensuring the periodic features of our model are
tied to the stellar rotation.
The corresponding posterior distributions for all the model parameters are showed in Fig. \ref{lnLikevsparam1planet_theta}. 
For the GP hyperparameter $w$ we get a symmetric distribution with all the values less than one, which is what is expected, as described, for example, in \cite{morales16}. Moreover, the distribution of the GP hyperparameter $\lambda$ has a median of 311 days and is nearly symmetric (excluding the points with $\lambda$<200 days that are relatively few compared to the whole sample, as indicated by the value of the 16th percentile). This long evolutionary timescale appears particularly interesting when compared with photometric results. 
\cite{wargelin16} show the evidence
for the rotational phasing to remain remarkably constant over the $\sim$15 years timespan at a fixed $P_{\rm rot}$=83.1 days, and for the modulation amplitude not to change significantly over two years, which could be indicative of the presence of persistent active longitudes on the stellar photosphere. For instance, the light curve corresponding to epochs between 2010 and 2012, which overlap with the timespan of the RV data, can be well described by a sinusoidal function with constant semi-amplitude and phase (see Fig. 3 in \citealt{wargelin16}), suggesting that the photospheric structures responsible for the correlations observed in the RV residuals could indeed have a decay timescale of the order of approximately one year.
The best-fit estimates for all the model parameters are showed in Table \ref{percentilesoneplan}, along with values for the planet orbital solution reported by AE16. The residuals of our global model have an r.m.s. of 1.3 $\ms$. In Fig. \ref{foldedrv1} we show the RV residuals, after removing the stellar signal described by the GP, folded at our best-fit estimate of the orbital period (Fig. \ref{foldedrv} shows the same plot for each dataset taken separately). It can be seen that a circular model well describes the data, in agreement with our finding that the eccentricity is consistent with zero within $\sim$1.4$\sigma$, whilst usually at least a 2.5$\sigma$ level is required for claiming a significant eccentricity. Fig. \ref{noisemodel} shows the RV residuals, after subtracting the Keplerian solution, 
with our
best-fit model for the correlated stellar noise superposed.
We note that the model describes the second
half of the HARPS-PRD dataset particularly well.

\begin{figure}[!htbp]
   \includegraphics[width=9.5cm]{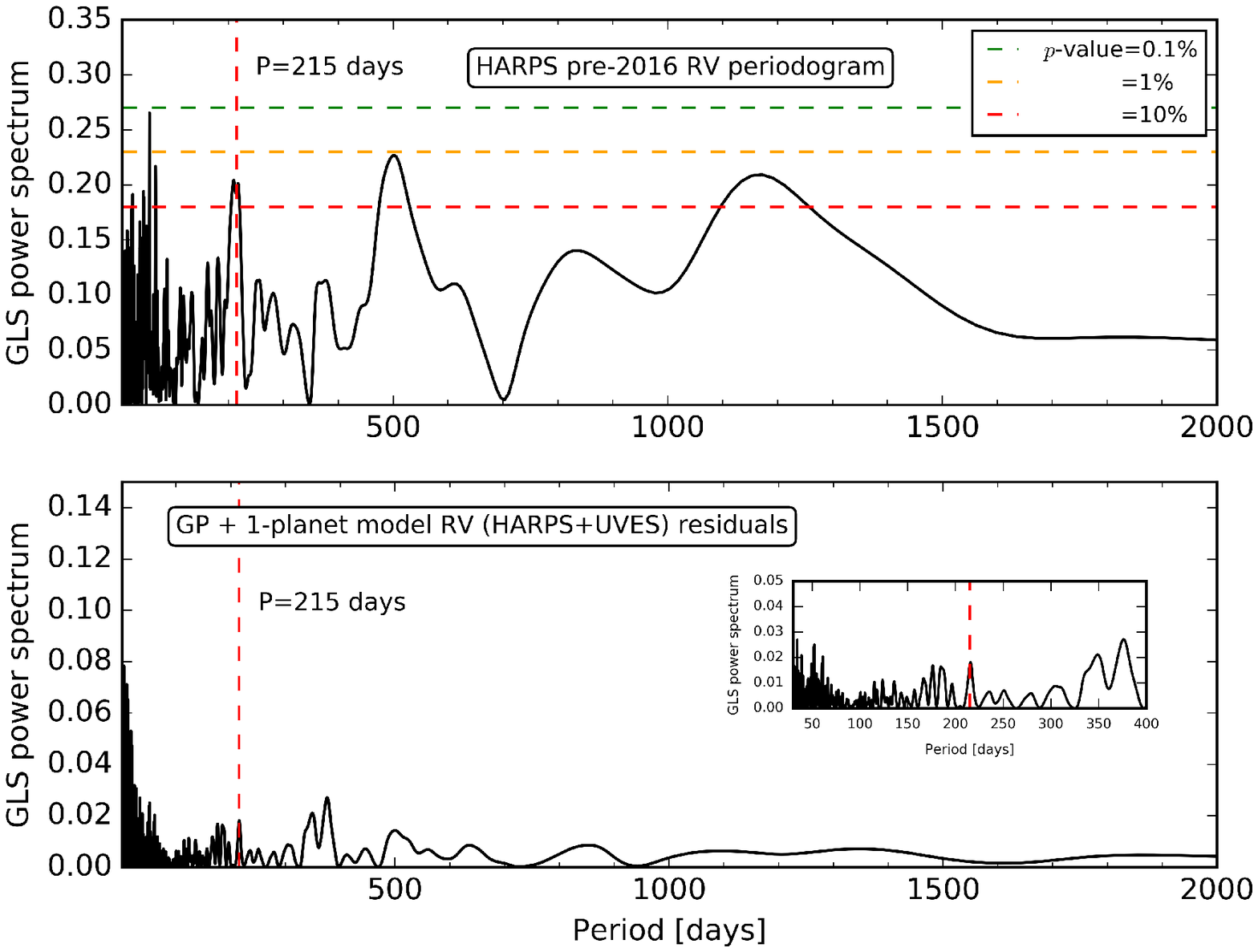}
   \caption{\textit{Upper plot}: GLS periodogram of the HARPS pre-2016 RV data. The vertical red dashed line marks P=215 days, the period found by AE16 in their likelihood-ratio periodogram, for which they could not give a physical interpretation. The \textit{p}-value levels have been derived through a bootstrap analysis.
   \textit{Lower plot}:  GLS periodogram of the complete RV datates residuals after subtracting the GP plus one-planet model. The inset plot shows the same periodogram around P=215 days, for more clarity.}
              \label{glsperiodogram1planet}%
\end{figure}

AE16 show (extended data Figure 2 therein) that their likelihood-ratio periodogram of the HARPS pre-2016 RV dataset has the highest peak at P=215 days (\textit{p}-value=1$\%$), but they could not conclude about the nature of this signal. To investigate in more detail the possible cause of its existence, we analyse the same dataset with the generalised Lomb-Scargle algorithm (GLS; \citealt{zech09}). We find that the highest power density peak appears at P$\sim$57 days (upper plot in Fig. \ref{glsperiodogram1planet}), while at P=215 days there is no significant power (\textit{p}-value>1$\%$, estimated through a bootstrap with re-sampling analysis on 10\,000 fake datasets). 
Then, we run GLS on the RV residuals of our global one-planet plus noise model. We do not find power excess in the periodogram at low frequencies (lower plot in Fig. \ref{glsperiodogram1planet}), indicating that our correlated noise model has suppressed any long-term modulation in the data. 
The impact of the GP model on these
residuals is not trivial to be characterised, and could prevent the
detection of additional planetary signals in the periodogram. Therefore,
we explored an additional two-planet plus stellar activity model to test
whether this signal is more effectively modelled as a planet or stellar activity.


\begin{table}[!htbp]
  \caption[]{Percentiles of the distributions in Fig. \ref{lnLikevsparam1planet_theta}, referred to the cut for $\theta$ in the range 85-89 days, compared with values found by AE16. They represent our adopted solution for the one-planet model. Our analysis makes use of 13 parameters as well as three values $\gamma$s for the offsets, while AE16 employed 26 parameters.}
         \label{percentilesoneplan}
         \scriptsize
   \begin{tabular}{c c c}
            \hline
            \noalign{\smallskip}
            Jump parameter     &  \multicolumn{2}{c}{Value}  \\
                               & [this work] & [Anglada-Escud\'{e} et al. 2016] \\
            \noalign{\smallskip}
            \hline
            \noalign{\smallskip}
            $h$ [\ms] & 1.91$^{+0.30}_{-0.23}$ \\
            \noalign{\smallskip}
            $\lambda$ [days] & 311$^{+71}_{-54}$  \\
            \noalign{\smallskip}
            $w$ & 0.34$^{+0.07}_{-0.06}$\\
            \noalign{\smallskip}
            $\theta$ [days] & 87.1$^{+0.9}_{-0.7}$ \\
            \noalign{\smallskip}
            \hline
            \noalign{\smallskip}
            $\gamma_{\rm HARPS_{pre-2016}}$ [\ms] & 0.92$\pm$0.56 \\
            \noalign{\smallskip}
            $\gamma_{\rm HARPS_{PRD}}$ [\ms] & -0.58$^{+0.94}_{-0.85}$  \\
            \noalign{\smallskip}
            $\gamma_{\rm UVES}$ [\ms] & -0.48$^{+0.39}_{-0.52}$ \\
            \noalign{\smallskip}
            \hline
            \noalign{\smallskip}
            $K_{\rm 1}$ [\ms] & 1.48$^{+0.13}_{-0.12}$ & 1.38$\pm0.21$ \\
            \noalign{\smallskip}
            $P_{\rm1}$ [days] & 11.1855$^{+0.0007}_{-0.0006}$ & 11.186$^{+0.001}_{-0.002}$\\
            \noalign{\smallskip}
            $T_{\rm c,1}$ [JD-2\,400\,000] & 57383.71$^{+0.24}_{-0.21}$ \\
            \noalign{\smallskip}
            $e_{\rm 1}$ & 0.17$^{+0.21}_{-0.12}$ & <0.35\\
            \noalign{\smallskip}
            $\omega_{\rm 1} [rad]$ & -2.49$^{+1.49}_{-0.35}$ & 5.41 (unconstrained) \\
            \noalign{\smallskip}
            \hline
            \noalign{\smallskip}
            $dV_{\rm r}/dt$ ($\cdot10^{-4}$) [\ms\,day$^{-1}$] & 3.2$^{+2.1}_{-3.0}$ & 2.3$\pm$8.4\\
            \noalign{\smallskip}
            $\sigma_{\rm jit, HARPS_{pre-2016}}$ [\ms] & 1.11$\pm$0.10 & 1.76$^{+0.60}_{-0.54}$\\
            \noalign{\smallskip}
            $\sigma_{\rm jit, HARPS_{PRD}}$ [\ms] & 0.63$^{+0.17}_{-0.19}$ & 1.14$^{+0.70}_{-0.57}$\\
            \noalign{\smallskip}
            $\sigma_{\rm jit, UVES}$ [\ms] & 0.71$^{+0.20}_{-0.18}$ & 1.69$^{+0.64}_{-0.47}$\\
            \noalign{\smallskip}
            \hline
            \noalign{\smallskip}
            Minimum mass, $m_{\rm p}\sini$ ($\mearth$)\tablefootmark{a} & 1.21$\pm0.16$ & 1.27$^{+0.19}_{-0.17}$\\
            \noalign{\smallskip}
            Orbital semi-major axis, $a$ (AU)\tablefootmark{a} & 0.048$\pm$0.002 & 0.0485$^{+0.0041}_{-0.0051}$\\ 
            \noalign{\smallskip}
            \hline
     \end{tabular}    
    \tablefoot{\tablefoottext{a}{Derived quantities from the posterior distributions. We assumed $M_{\rm s}$=0.120$\pm$0.015 $\Msun$ as the mass of Proxima, and we used the following equations (assuming  $M_{\rm s}+m_{\rm p} \cong M_{\rm s}$): $m_{\rm p}\sini \cong$ ($K_{\rm 1} \cdot M_{\rm s}^{\frac{2}{3}} \cdot \sqrt{1-e^{2}} \cdot P_{\rm 1}^{\frac{1}{3}}) / (2\pi G)^{\frac{1}{3}}$; $a \cong [(M_{\rm s}\cdot G)^{\frac{1}{3}}\cdot P_{\rm 1}^{\frac{2}{3}}]/(2\pi)^{\frac{2}{3}} $, where $G$ is the gravitational constant. }}
   \end{table}

\section{Two-planet model}
\label{sec:twopl}
\begin{figure*}
   \caption{Marginal posterior distributions for the two-planet model (circular option).}
   \includegraphics[width=19cm]{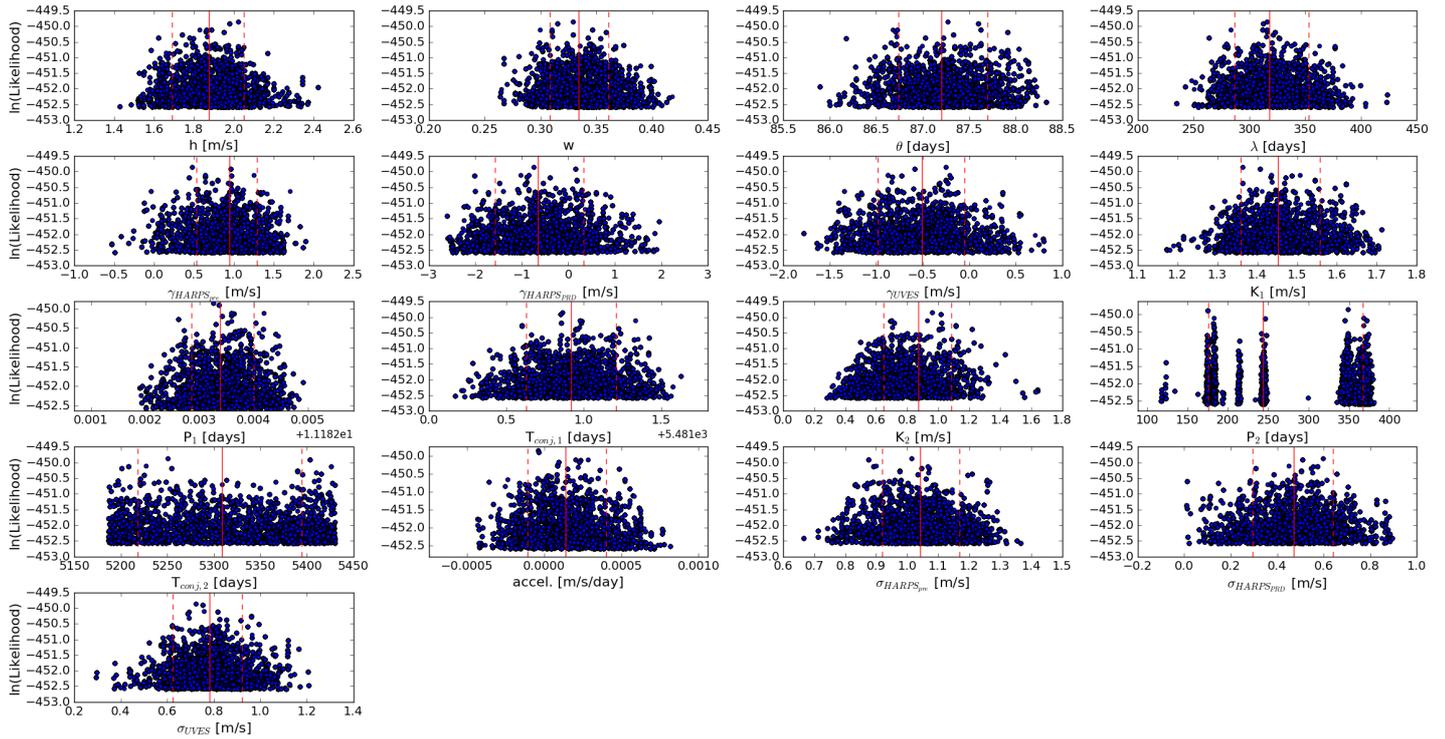}
              \label{postdist2planetmodel}%
\end{figure*}

We tested the two-planet scenario using Gaussian priors for all the GP hyperparameters, except for \textit{h}, based on the physically reliable results obtained for the stellar noise representation for the one-planet model. In modelling the RVs we ignored the mutual gravitational perturbations between the two planets. Taking further advantage of the one-planet model results to speed up the analysis, for Proxima\,b we fixed the upper limit for RV semi-amplitude to 3 \ms, that of the orbital period to 20 days and the eccentricity to zero. We explore the parameter space for the orbital period of the second Keplerian starting from 100 days, on the basis that no clear modulation with a period equal or less than that is observed in the HARPS PRD RV dataset (see AE16). 
We ran both a model where the orbit of the outer planet is assumed circular, and one where the eccentricity is treated as a free parameter.
The list of the adopted priors is showed in Table \ref{prior2plan}.
The marginal posterior distributions for the simpler circular model are showed in Fig. \ref{postdist2planetmodel}. 
For this case, we note that the orbital period of the outer planet appears to be multi-modal, and the semi-amplitude of the Keplerian is below the average value of the internal RV uncertainties ($\left\langle\sigma_{\rm RV}\right\rangle$=0.94 $\ms$). Moreover, the r.m.s. of the residuals remains the same between the one and two-planet model ($1.3$ and $1.2 \ms$, respectively). These evidences alone do not favour the existence of a second Keplerian signal in the data. A similar conclusion can be drawn for the model with the eccentricity as free parameter. Here, the eccentricity appears to be  insignificant within $2\sigma$.



\begin{table}
  \caption[]{Prior probability distributions for the two-planet model parameters. The orbit of the inner planet is assumed circular.}
         \label{prior2plan}
         \scriptsize
   \begin{tabular}{l l l l l l}
    
            \hline
            \noalign{\smallskip}
            Jump parameter     &  Prior & Lower & Upper & Mean & $\sigma$ \\
                               &        & bound & bound & & \\
            \noalign{\smallskip}
            \hline
            \noalign{\smallskip}
            h [\ms] & Uniform & 0 & 10 & & \\
            $\lambda$ [days] & Gaussian & - & - & 311 & 71 \\
            $w$ & Gaussian & - & - & 0.34 & 0.07 \\
            $\theta$ [days] & Gaussian & - & - & 87.1 & 0.9 \\
            $K_{\rm 1}$ [\ms] & Uniform & 1 & 3 & & \\
            $\ln P_{\rm1}$ [days] & Uniform & 0 (P$_{1}$=1) & $\ln20$ & & \\
            $T_{\rm c,1}$ [JD-2\,400\,000] & Uniform & 51500 & 57500 & & \\
            $K_{\rm 2}$ [\ms] & Uniform & 0 & 100 & & \\
            $\ln P_{\rm2}$ [days] & Uniform & 4.6 (P$_{2}$=100) & $\ln6000$ & &\\
            $T_{\rm c,2}$ [JD-2\,400\,000] & Uniform & 57400 & 63400 & &  \\
            $C_{\rm 2}=\sqrt[]{e_{\rm 2}}\cdot\cos\omega_{\rm 2}$ & Uniform & -1 & 1 & & \\
            $S_{\rm 2}=\sqrt[]{e_{\rm 2}}\cdot\sin\omega_{\rm 2}$ & Uniform & -1 & 1 & & \\
            $e_{\rm 2}$ (=$C_{\rm 2}^{2}$+$S_{\rm 2}^{2}$) & 3.1$\cdot(1-e_{\rm 2})^{2.1}$ & 0 & 1 & & \\
            $\omega_{\rm 2}$ [rad] & Uniform & 0 & 2$\pi$\\
            $dV_{\rm r}/dt$ [\ms\,day$^{-1}$] & Uniform & -0.001 & +0.001 & & \\
            $\gamma_{\rm HARPS_{pre-2016}}$ [\ms] & Uniform & -3 & +3 & & \\
            $\gamma_{\rm HARPS_{PRD}}$ [\ms] & Uniform & -3 & +3 & & \\
            $\gamma_{\rm UVES}$ [\ms] & Uniform & -3 & +3 & & \\
            $\sigma_{\rm jit, HARPS_{pre-2016}}$ [\ms] & Uniform & 0 & 5 & & \\
            $\sigma_{\rm jit, HARPS_{PRD}}$ [\ms] & Uniform & 0 & 5 & & \\
            $\sigma_{\rm jit, UVES}$ [\ms] & Uniform & 0 & 5 & & \\
            \noalign{\smallskip}
            \hline           
     \end{tabular}    
   \end{table}

\begin{table}[h]
  \caption[]{Two-planet model parameter estimates, obtained by fixing the eccentricity of the second Keplerian to zero, and also keeping it as a free parameter.}
         \label{posterior2plan}
         \centering
         \scriptsize
   \begin{tabular}{ccc}
            \hline
            \noalign{\smallskip}
            Jump parameter     &  \multicolumn{2}{c}{Value} \\
                               &  circular orbit & eccentric orbit  \\ 
            \noalign{\smallskip}
            \hline
            \noalign{\smallskip}
            h [\ms] & 1.87$\pm$0.18 & 1.81$^{+0.18}_{-0.07}$ \\
            \noalign{\smallskip}
            $\lambda$ [days] & 317$^{+35}_{-21}$ & 337$^{+25}_{-34}$ \\
            \noalign{\smallskip}
            $w$ & 0.33$\pm$0.03 & 0.33$^{+0.03}_{-0.04}$ \\
            \noalign{\smallskip}
            $\theta$ [days] & 87.2$\pm$0.5 & 87.1$^{+0.5}_{-0.7}$\\
            \noalign{\smallskip}
            \hline
            \noalign{\smallskip}
            $\gamma_{\rm HARPS_{pre-2016}}$ [\ms] & 0.95$^{+0.34}_{-0.41}$ & 0.88$^{+0.38}_{-0.36}$ \\
            \noalign{\smallskip}
            $\gamma_{\rm HARPS_{PRD}}$ [\ms] & -0.65$^{+0.97}_{-0.92}$ & 0.35$^{+0.46}_{-1.04}$ \\
            \noalign{\smallskip}
            $\gamma_{\rm UVES}$ [\ms] & 0.51$\pm$0.47 & -0.70$^{+0.63}_{-0.38}$ \\
            \noalign{\smallskip}
            \hline
            \noalign{\smallskip}
            $K_{\rm 1}$ [\ms] & 1.45$^{+0.10}_{-0.09}$ & 1.45$^{+0.09}_{-0.07}$ \\
            \noalign{\smallskip}
            $P_{\rm1}$ [days] & 11.1854$^{+0.0006}_{-0.0005}$ & 11.1854$^{+0.0005}_{-0.0007}$ \\
            \noalign{\smallskip}
            $T_{\rm c,1}$ [JD-2\,400\,000] &  57383.57$\pm$0.19 &  57383.57.59$\pm$0.20 \\
            \noalign{\smallskip}
            \hline
            \noalign{\smallskip}
            $K_{\rm 2}$ [\ms] & 0.87$\pm$0.22 & 1.17$^{+0.74}_{-0.30}$ \\
            \noalign{\smallskip}
            $P_{\rm2}$ [days] & 243$^{+123}_{-68}$ & 344$^{+22}_{-99}$ \\
            \noalign{\smallskip}
            $T_{\rm c,2}$ [JD-2\,400\,000] & 60308$^{+86}_{-91}$ & 60704$^{+125}_{-71}$  \\
            \noalign{\smallskip}
            $e_{\rm 2}$ & 0 & 0.46$\pm0.26$ \\
            \noalign{\smallskip}
            $\omega_{\rm 2}$ [rad] & 0 & -0.15$^{+1.1}_{-2.3}$ \\
            \noalign{\smallskip}
            \hline
            \noalign{\smallskip}
            $dV_{\rm r}/dt$ [\ms\,day$^{-1}$] & 1.4$\pm2.6\cdot10^{-4}$ & 3$^{+21}_{-17}(\cdot10^{-5})$ \\
            \noalign{\smallskip}
            $\sigma_{\rm jit, HARPS_{pre-2016}}$ [\ms] & 1.04$\pm$0.13 & 0.98$^{+0.17}_{-0.08}$ \\
            \noalign{\smallskip}
            $\sigma_{\rm jit, HARPS_{PRD}}$ [\ms] & 0.45$\pm$0.17 & 0.56$^{+0.12}_{-0.23}$ \\
            \noalign{\smallskip}
            $\sigma_{\rm jit, UVES}$ [\ms] & 0.78$^{+0.14}_{-0.16}$ & 0.67$^{+0.18}_{-0.16}$ \\
            \noalign{\smallskip}
            \hline
     \end{tabular}    
   \end{table}

\begin{table}[h]
  \caption[]{Bayesian evidences for the tested models.}
         \label{bayesev}
         \scriptsize
   \begin{tabular}{l c c}
            \hline
            \noalign{\smallskip}
            Model     &  \multicolumn{2}{c}{$\ln\mathcal{Z}$} \\
                      &  [CJ14] & [Perr14] \\
            \noalign{\smallskip}
            \hline
            \noalign{\smallskip}
            one-planet & -444.6$\pm$0.3 & -440.6$\pm$0.3 \\
            two-planet (circular)& -455.7$\pm$0.2 & -458.67$\pm$0.03 \\
            two-planet (free eccentricity)& -454.6$\pm$0.6 & -457.25$\pm$0.05\\
            \hline
            \noalign{\smallskip}
            $\Delta(\ln\mathcal{Z}$), one-planet minus two-planet (circular) & +11 & +18 \\
            $\Delta(\ln\mathcal{Z}$), one-planet minus two-planet (free eccentricity) & +10 & $\sim$+17 \\
            \noalign{\smallskip}
            \hline
     \end{tabular}    
 \end{table}

\section{Summary and conclusions}
\label{sec:concl}
We used a Gaussian process framework to allay the stellar correlated noise in the radial velocity timeseries of Proxima Centauri published by \cite{anglada16}. In our fitting procedure we considered only radial velocity measurements, without including other ancillary data. We then compared the outputs of our procedure with those coming from photometry, to provide a consistent physical interpretation of the results. We adopted a quasi-periodic kernel to describe the correlated noise: this is a widely used function that represents the covariance between measurements at different epochs in terms of parameters that can be related to some physical properties of the star. The philosophy behind our approach - but not the analysis tools - is similar to that of \cite{faria16}, which used GPs to model the stellar noise in RVs of the host-star CoRoT-7 and were able to retrieve planets CoRoT-7\,b and CoRoT-7\,c without analysing any other additional dataset.

In our study, we confirm the detection of a coherent signal well described by a Keplerian orbit equation that can be attributed to the planet Proxima\,b. We find system parameters to be in good agreement with those of \cite{anglada16}. We note that our estimates of the uncorrelated jitter terms for each independent RV dataset are lower than in \cite{anglada16} (Table \ref{percentilesoneplan}), suggesting that our model removes the stellar noise more efficiently. 

As a major outcome, the correlated noise has indeed a periodic component modulated by the stellar rotation period, and we can describe it through a simple GP-based regression. The best-fit estimates of all the hyperparameters assume realistic physical values, as the stellar rotation period or the typical lifetime of active regions, because we have a measured or deduced counterpart from an independent photometric dataset. Therefore our model is physically trustworthy as well as simpler, because it makes use of fewer free parameters. This is likely the reason because we find smaller errors for the Doppler semi-amplitude and the orbital period.
We note that \cite{anglada16} do not mention the existence in their RV periodograms of a significant signal with a frequency close to the stellar rotation period, but in light of our results the rotation of Proxima could be considered the main cause of the clear trend visible in the high-sampling HARPS-PRD dataset.

Once constrained the stellar noise on the results found for the one-planet model, we tested the presence of a second companion by adding a Keplerian to the model. Table \ref{bayesev} shows the Bayesian evidences for the considered models from two different estimators. Both strongly favour the one-planet scenario, with $\Delta(\ln\mathcal{Z}$) between 10 and 18. Our analysis thus shows that, with our proposed treatment of the stellar noise, the present RV dataset does not show unambiguous evidence for the existence of an additional, low-mass planetary companion to Proxima on an outer orbit with an orbital period of between 100 to 6\,000 days.

\begin{acknowledgements}
     This work was conceived and partially carried out during the \textit{Approaching the Stellar Astrophysical Limits of Exoplanet Detection: Getting to 10 cm/s} workshop held at the Aspen Center for Physics on Sept. 2016. The Aspen Center for Physics is supported by National Science Foundation grant PHY-1066293. MD was granted by the Simons Foundation to attend the workshop, and he acknowledges funding from Progetto Premiale INAF “Way to Other
     Worlds” (WoW), decreto 35/2014. FDS acknowledges the Swedish Research Council International Postdoc fellowship for support. We thank A. Sozzetti, A.S.Bonomo and R. Haywood for valuable discussions, and the referee for useful comments. We finally thank M. Boldi and C. De Sica for some pleasant hours spent in Aspen, twenty years in the past.
\end{acknowledgements}

\bibliographystyle{aa} 
\bibliography{proxima.bib} 


\Online

\end{document}